\documentclass[12pt]{article}

\def\rdots{\mathinner{\mkern1mu\raise1pt\vbox{\kern1pt\hbox{.}}\mkern2mu
   \raise4pt\hbox{.}\mkern2mu\raise7pt\hbox{.}\mkern1mu}}
\newcommand{\Z}{{\rm Z\kern-.35em Z}}
\newcommand{\bP}{{\rm I\kern-.15em P}}
\newcommand{\Q}{\kern.3em\rule{.07em}{.65em}\kern-.3em{\rm Q}}
\newcommand{\R}{{\rm I\kern-.15em R}}
\newcommand{\h}{{\rm I\kern-.15em H}}
\newcommand{\C}{\kern.3em\rule{.07em}{.65em}\kern-.3em{\rm C}}
\newcommand{\T}{{\rm T\kern-.35em T}}

\newcommand{\be}{\begin{equation}}
\newcommand{\ee}{\end{equation}}
\newcommand{\bea}{\begin{eqnarray}}
\newcommand{\eea}{\end{eqnarray}}

\newcommand{{\ba}}{\bf a}

\begin{document}

\openup 1.5\jot

\centerline{A Note on Singular Instantons}

\vspace{.25in}
\centerline{Paul Federbush}
\centerline{Department of Mathematics}
\centerline{University of Michigan}
\centerline{Ann Arbor, MI 48109-1109}
\centerline{(pfed@math.lsa.umich.edu)}

\vspace{.50in}

\centerline{Abstract}
\ \ \ \ \ We point out the existence of some singular, radial, spin-0 instantons for curvature-quadratic gravity theories.  They are complex.

\vspace{.25in}

\centerline{-----------------------------------------------}

We consider Euclidean gravity theories described by the Lagrangian density
\be	\sqrt{g} \Big( \alpha R_{ik} R^{ik} + \beta R^2 \Big) .	\ee
For these theories there are solutions of the Euler-Lagrange equations of the form
\be	g_{\mu \nu}(x) = \delta_{\mu \nu}  \Big(r^2\Big)^\varepsilon	\ee
for $\varepsilon = -2, \ -1 \pm i/\sqrt{3}$.  Here
\be	r^2 = \sum^4_{i=1} x^2_i \ .	\ee
For $\varepsilon = -2$ the instanton is also a (well-known, trivial) instanton of the Einstein action.  For the two complex values of $\varepsilon$ this is not true, and the values of the metric are complex.  For $\beta = - \frac 1 3 \; \alpha$, any $g_{\
mu \nu}$ of the form
\be	g_{\mu \nu}(x) = \delta_{\mu \nu} \; f(x)	\ee
satisfies the Euler-Lagrange equations; one is dealing with a conformal gravity theory.
\end{document}